\documentclass{article}
\usepackage{spconf,hyperref}
\usepackage{amsmath,amssymb,amsfonts}
\usepackage{algorithmic}
\usepackage{graphicx}
\usepackage{textcomp}
\usepackage{xcolor}
\usepackage{balance}
\usepackage{cite}
\usepackage{bm}
\usepackage{multirow}
\usepackage{pgfplots}
\usepackage{pgfplotstable}
\pgfplotsset{compat=1.8}
\usetikzlibrary{plotmarks, calc}

\title{Phase-Retrieval-Based Physics-Informed Neural Networks\\For Acoustic Magnitude Field Reconstruction}

\name{Karl Schrader$^{1,2}$%
\qquad Shoichi Koyama$^{1}$%
\qquad Tomohiko 
Nakamura$^{3}$ \qquad Mirco Pezzoli$^{4}$
}
  
\address{
$^{1}$\textit{National Institute of Informatics}, 
Tokyo, Japan, \
$^{2}$ \textit{Saarland University}, 
Saarbr\"ucken, Germany \\
$^{3}$ \textit{National Institute of Advanced Industrial Science and Technology}, 
Tokyo, Japan\\
$^{4}$ \textit{Politecnico di Milano}, 
Milan, Italy\\
\thanks{This work was supported by JSPS KAKENHI under Grant Number 23K24864, JST FOREST Program under Grant Number JPMJFR216M, "Strategic Programs" grant from ROIS (Research Organization of Information and Systems), and the Deutsche Forschungsgemeinschaft (DFG, German Research Foundation) under project number 471607914.}
}

\begin{document}
\ninept
\maketitle
\begin{abstract}
We propose a method for estimating the magnitude distribution of an acoustic field from spatially sparse magnitude measurements. Such a method is useful when phase measurements are unreliable or inaccessible. Physics-informed neural networks (PINNs) have shown promise for sound field estimation by incorporating constraints derived from governing partial differential equations (PDEs) into neural networks. However, they do not extend to settings where phase measurements are unavailable, as the loss function based on the governing PDE relies on phase information. To remedy this, we propose a phase-retrieval-based PINN for magnitude field estimation. By representing the magnitude and phase distributions with separate networks, the PDE loss can be computed based on the reconstructed complex amplitude. We demonstrate the effectiveness of our phase-retrieval-based PINN through experimental evaluation.
\end{abstract}
\section{Introduction}
Sound field estimation/reconstruction/interpolation aims to estimate a spatial distribution of an acoustic field based on a discrete set of sensor observations. It is one of the fundamental techniques in acoustic signal processing and machine learning that can be applied to various downstream tasks, such as acoustic imaging~\cite{Maynard:JASA1985}, room acoustic analysis~\cite{Park:JASA_J_2005}, and spatial audio reproduction~\cite{Poletti:J_AES_2005}. 

Sound field estimation methods are generally targeted at complex-valued amplitude distribution in the frequency domain or real-valued amplitude distribution in the time domain, and assume that they can be observed at discrete positions of multiple sensors. One of the most widely used techniques is the basis-expansion-based method, which is based on the expansion representation of the acoustic field by plane wave functions, spherical wave functions, or equivalent sources \cite{Poletti:J_AES_2005,Park:JASA_J_2005,Koyama:IEEE_J_JSTSP2019}.
The basis-expansion-based methods have been generalized as kernel-regression-based methods, which constrain the estimated function to satisfy the Helmholtz equation by properly-defined kernel functions~\cite{Ueno:IEEE_J_SP2021}. 
Recently, neural-network-based (NN-based) methods have attracted attention due to their high flexibility and representational power~\cite{Lluis:JASA2020,Luo:NIPS2022,Ribeiro:IEEE_ACM_J_ASLP2024,Olivieri:EURASIP2024,Koyama:ForumAcusticum2025}.
Among the NN-based methods, the physics-informed neural network 
(PINN)~\cite{Karniadakis:NatRevPhys2021,Koyama:IEEE_M_SP2025} is considered one of the promising techniques because it uses an explicit constraint on physical properties, defined as the deviation of the NN output from the governing partial differential equations (PDEs), i.e., wave or Helmholtz equation~\cite{Raissi:CompPhys2019,BorrelJensen:PNAS2024}. 
This constraint allows the PINN to avoid overfitting by penalizing functions that do not adhere to the governing PDE. 

In contrast to many sound field estimation methods, the target of this study is the acoustic magnitude distribution, i.e., the absolute value of the complex-valued amplitude (or pressure), of the sound field in the frequency domain with its discrete measurements. 
The estimation techniques for acoustic magnitude distribution will be useful when the signals of each sensor are not synchronized. For example, in ad-hoc microphones, each sensor device usually operates independently~\cite{Cobos:WCMC2017}, and when measuring the directivity of vibrating bodies, such as musical instruments, sequential recording of sound radiation is frequently used~\cite{Ackermann:EURASIP_JASM2021}. 
However, unlike the methods using the complex-valued amplitudes, it is difficult for the magnitude distribution estimation to incorporate the physical properties of the sound field because the governing PDE cannot be computed without the phase. Therefore, the magnitude field estimation has typically relied on purely data-driven techniques. 

We propose a magnitude field estimation method that incorporates phase retrieval and enables the use of a Helmholtz-equation-based loss even without access to phase measurements.
Specifically, our NN jointly estimates magnitude and phase distributions so that the measured magnitudes are matched at sensor positions while the deviation from the Helmholtz equation is minimized.
The effectiveness of our phase-retrieval-based PINN is evaluated through numerical experiments in simulated rooms.

\section{Problem Statement and Prior Works}

\subsection{Acoustic Magnitude Field Estimation}
We consider an arbitrary room with target region $\Omega \subset
\mathbb{R}^3$. Inside the target region, the magnitude of the acoustic pressure
$\vert u(\bm x, \omega)\vert\!: \Omega \times \mathbb{R} \to \mathbb{R}$ is 
known for a set of $M$ measurement locations $\{\bm x_m^{\mathrm{(m)}}\}_{m=1}^M 
\in \Omega$, and a fixed angular frequency $\omega$. The goal is to infer the magnitude of the acoustic pressure at a set of $N$ test locations 
$\{\bm x_n^{\mathrm{(t)}}\}_{n=1}^N \in \Omega$ that are different from the measured ones. In the following, we will omit $\omega$ for notational simplicity. Note that this problem differs from the usual estimation of the complex-valued acoustic pressure in that only the magnitude is known, but not the 
phase. 

\subsection{Related Works}

Most previous works consider the interpolation of the complex-valued pressure 
field instead of just its magnitude. 
The most widely used approach in the pressure interpolation is the 
basis-expansion-based method, which is based on the expansion of the pressure 
distribution into plane waves, spherical wave functions, or point 
sources~\cite{Colton:InvAcoust_2013,Ueno:FnT_SP2025}. This technique is 
generalized as kernel regression with a constraint of the governing PDE, 
which can be interpreted as an infinite-dimensional basis 
expansion~\cite{Ueno:IEEE_J_SP2021}. 

Recently, many NN-based methods for sound field estimation and head-related 
transfer function (HRTF) interpolation, which is closely related to sound field 
estimation, have been 
proposed~\cite{Lluis:JASA2020,sanchez2025towards, Luo:NIPS2022,Su:NeurIPS2022,Ribeiro:IEEE_ACM_J_ASLP2024,Olivieri:EURASIP2024,Ito:IWAENC2022,Zhang:ICASSP2023,Koyama:ForumAcusticum2025}.
 Several of these, particularly in HRTF interpolation, target the interpolation 
of magnitude distributions~\cite{Lluis:JASA2020,Ito:IWAENC2022,Zhang:ICASSP2023,Koyama:ForumAcusticum2025}.
 PINN enables the combination of the high representational power of NNs with physical prior knowledge~\cite{Karniadakis:NatRevPhys2021,Koyama:IEEE_M_SP2025}.
 Based on the implicit neural representation (or neural field: NF)~\cite{Sitzmann:NeurIPS2020}, PINN represents the spatial distribution using a NN and trains it using a loss function that induces the output to satisfy the governing PDE, which is called PDE loss~\cite{Raissi:CompPhys2019}. However, in PINN, since the PDE loss is computed via automatic differentiation from the NF output~\cite{Baydin:JMLR2018}, 
the target is limited to the complex-valued amplitude in the frequency domain. 
Therefore, the same methods used for interpolating general real-valued functions 
have been generally applied to estimate magnitude distributions. 

Meanwhile, even when estimating the magnitude distribution, the original 
complex-valued amplitude distribution should be constrained to the space spanned 
by functions satisfying the governing PDEs. Consequently, the function space 
spanned by the magnitude distribution is also considered to be constrained. 

\begin{figure}[t!]
\center{\includegraphics[width=0.85\columnwidth]{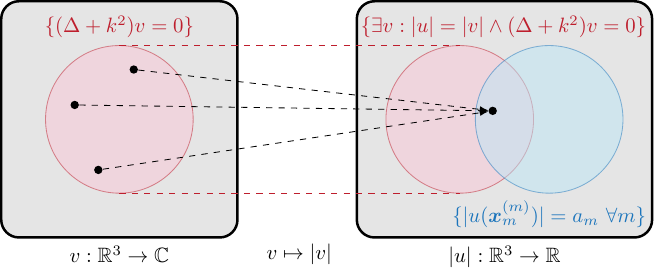}}
\caption{Relationship of pressure fields and magnitude fields. 
The goal of our approach is to find a function which fulfills the two conditions, symbolised by the two circles.
}
\label{fig:motivation}
\end{figure}






\section{Phase-retrieval-based PINN for magnitude field estimation}

We begin with a short introduction of PINNs at the example of a complex-valued 
pressure field estimation, before describing how we extend the approach to be 
able to apply it to magnitude field estimation.

\begin{figure}[t!]
\includegraphics[width=\columnwidth]{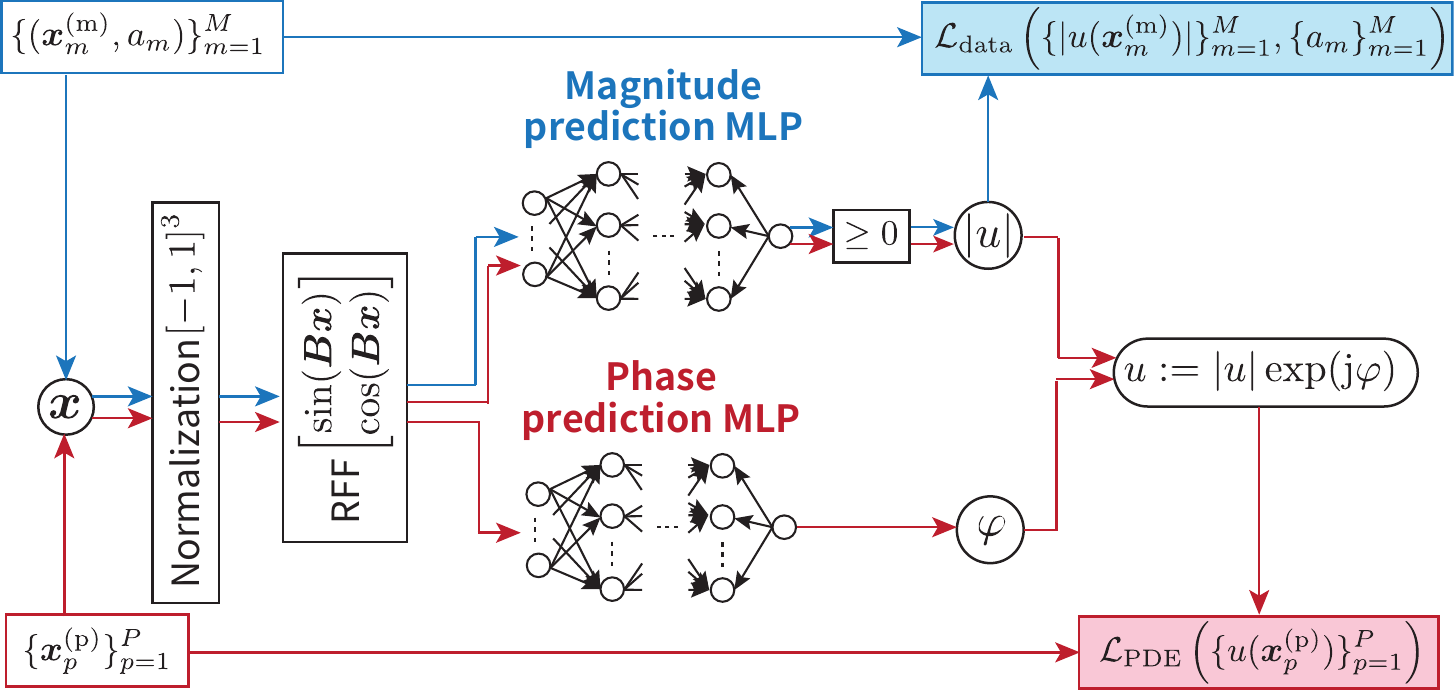}
\caption{Visualization of our network architecture. Blue arrows highlight the data flow of the magnitude dataset, and red arrows of the reconstructed complex-valued pressure dataset.}
\label{fig:architecture}
\end{figure}

\begin{figure}[tbp]
\center{
\includegraphics[clip,width=0.6\columnwidth]{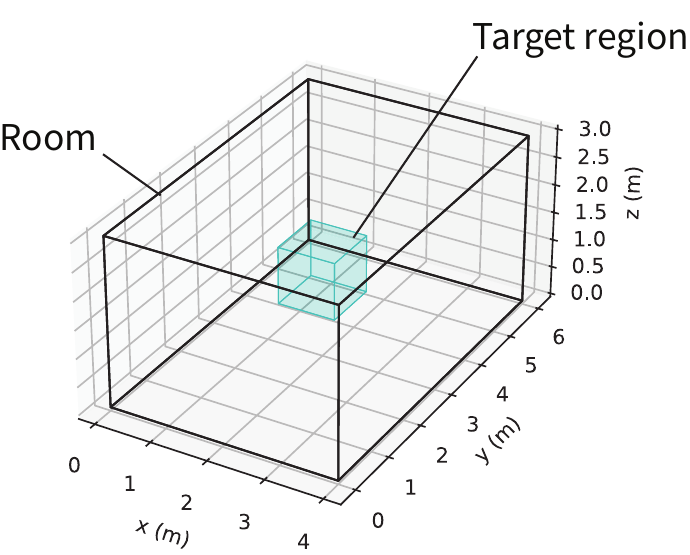}
}
\caption{Experimental setup. Geometry of the room and target region~$\Omega$.}
\label{fig:geometry}
\end{figure}

\begin{figure*}[t!]
\centering
\includegraphics[width=\linewidth]{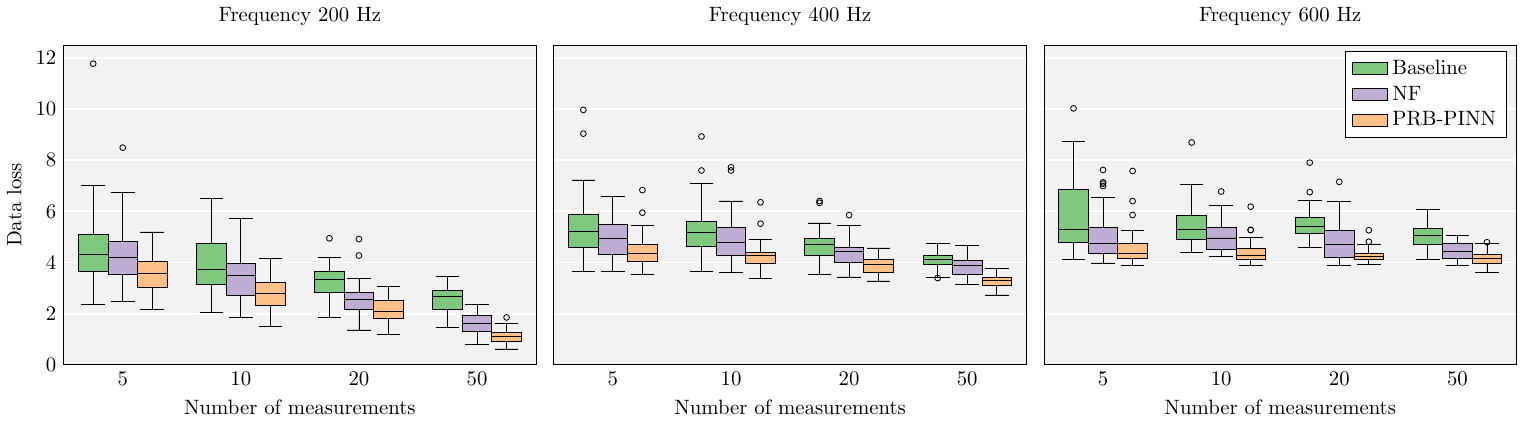}
\caption{Plot of the test set data loss \eqref{eq:data_loss} for different 
frequencies and numbers of sources in the training set. The use of the physics 
loss is beneficial in all cases.}
\label{fig:data_loss}
\end{figure*}

\subsection{PINN}
PINNs~\cite{Raissi:CompPhys2019} aim to solve initial value problems (IVP) by 
modeling the target function with a NF, and make use of automatic differentiation 
frameworks for the calculation of partial derivatives. 
The IVP that is considered for sound field estimation in the frequency domain is 
defined by the homogeneous Helmholtz equation 
\begin{equation}
(\Delta + k^2)u(\bm x)=0,\qquad u:\Omega\times\mathbb R \to \mathbb C,\ \bm x 
\in \Omega , \label{eq:helmholtz}
\end{equation}
where $k$ is the wave number, together with the initial condition
\begin{equation}
u(\bm x_m^{\mathrm{(m)}}) = s_m, \qquad m = 1, \dots, M\,.
\end{equation}
Here, $\{s_m\}_{m=1}^M$ are the pressure measurements at the locations $\{\bm{x}_m\}_{m=1}^M$.

PINNs model the sound field $u(\bm x)$ through a NF. Its input are spatial 
coordinates $\bm x \in \Omega$, and its output is the predicted sound field at 
that location. To ensure that this NN adheres to the initial condition and 
Helmholtz equation, two loss functions are defined. The data loss 
$\mathcal{L}_{\text{data}}$ is used to penalize deviations from the initial 
condition. The PDE loss $\mathcal L_{\text{PDE}}$ penalises deviation from the 
governing PDE. To that end, locations $\{\bm{x}_p^{\mathrm{(p)}}\}, p=1,\dots,P$, 
are sampled from the spatial domain $\Omega$. At those locations, the mean 
squared residual is considered:
\begin{equation}
\mathcal{L}_{\text{PDE}} = \frac1P\sum_{p=1}^{P} \left\vert(\Delta + k^2)
u(\bm x_p^{\mathrm{(p)}})\right\vert_2^2\,. \label{eq:physics_loss}
\end{equation}
A weighting of both losses is used for training:
\begin{equation}
\mathcal{L} = \lambda_{\text{data}} \mathcal{L}_{\text{data}} + 
\lambda_{\text{PDE}}\mathcal{L}_{\text{PDE}}\,.
\end{equation}
The weights $\lambda_{\text{data}}, \lambda_{\text{PDE}}>0$ can be 
chosen to prioritise one loss over the other. A technique to adapt the weighting 
parameters is also proposed~\cite{Xiang:NeuroComp2022}.

During training, the loss $\mathcal L$ is minimised through gradient descent. 
Partial derivatives inside the Helmholtz equation are calculated using automatic 
differentiation~\cite{Baydin:JMLR2018}, which saves the user from having to 
discretize them.
It is clear to see that when both losses are~$0$, the IVP is solved. While PINNs 
typically only reach this minimum approximately, they are still good 
approximations of the target function.

Notably, the setup described here cannot be directly applied to the magnitude 
field estimation. The Helmholtz equation~\eqref{eq:helmholtz} requires the 
complex-valued pressure field, which is not available in our setting. 
Additionally, no equivalent statement operating on the magnitude fields $\lvert u 
\rvert$ is known.

\subsection{Motivation of Our Approach}%

The key insight of our approach is that there has to be some underlying sound 
field $u$ which fulfills two conditions:
\begin{enumerate}
\item At the sensor locations $\bm x_m^{\mathrm{(m)}}$, its magnitude should 
match the observed one $\lvert s_m\rvert:=a_m$ ($m=1,\dots,M$).
\item The sound field $u$ adheres to the Helmholtz equation everywhere in the 
domain.
\end{enumerate}
This sound field cannot be uniquely determined, as the phase component can be 
periodically shifted. Additionally, for a low number of sensors $M$, there 
can be further degrees of freedom. Accordingly, a sound field $u$ fulfilling these conditions can still differ significantly from ground truth. We visualize this relationship in Fig.~\ref{fig:motivation}. However, as our experiments in Sec.~\ref{sec:experiments} will show, using the Helmholtz equation as a regularizer in this way still improves reconstruction quality.

For our model, we encode the two conditions mathematically:
\begin{enumerate}
\item \textbf{Data Condition:}
\begin{equation}
\lvert u(\bm x_m^{\mathrm{(m)}})\rvert = a_m\qquad m=1,\dots,M
\end{equation}
\item \textbf{Physics Condition:}
\begin{equation}
(\Delta + k^2)u(\bm x)=0\qquad\forall\bm x\in\Omega
\end{equation}
\end{enumerate}
In the following, we demonstrate how to train a PINN to find a sound field $u$ 
fulfilling these conditons.

\subsection{Proposed Architecture}

Similar to the PINN approach for sound field estimation, we predict a sound field 
$u$, and encode our two conditions through loss functions.
For the data condition, whereas the commonly used mean square error could be used, here we selected the mean average error of the logarithmically scaled magnitude as a perceptually-motivated data loss~\cite{Xie:AAA2010}
\begin{equation}
\mathcal{L}_{\text{data}} = \frac{1}{M}\sum_{m=1}^{M}\left|
20\log_{10}\left(\frac{a_m}{|u(\bm x_m^{\mathrm{(m)}})|}\right)\right|,
\label{eq:data_loss}
\end{equation}
which is equivalent to the \textit{log-spectral distance} for a single frequency.
For the physics condition, we penalize the squared 
residual~\eqref{eq:physics_loss}.

Our final network architecture is illustrated in Fig. \ref{fig:architecture}. 
At its core, it consists of two NFs with multilayer perceptrons (MLPs), one for magnitude estimation and one for phase prediction. As input, both receive a random Fourier feature (RFF)~\cite{Tancik:NeurIPS2020}, which maps the sensor position into high-dimensional features using sinusoids with the randomly sampled frequencies $\bm{B}$. 
This greatly improves the network's ability to handle higher-frequency variations 
in the output.

\section{Experiments} \label{sec:experiments}

\subsection{Experimental Setup}

\noindent\textbf{Dataset.} We train on a synthetic dataset generated with the image source method~\cite{Allen:JASA1979} using 
\verb|pyroomacoustics|~\cite{pyroomacoustics}. Our environment is a room of size $3~\mathrm{m}\times4~\mathrm{m}\times6~\mathrm{m}$ with a reverberation time $T_{60}$ of $200~\mathrm{ms}$ as shown in Fig.~\ref{fig:geometry}. 
Inside, a cube lattice of $33^3$ positions is placed in a unit cube ($1.0~\mathrm{m}^3$) at the origin as the target region~$\Omega$. Of those, $5$, $10$, $20$, or $50$ measurement locations are randomly selected, and the rest are used for testing.
$64$ sources are placed randomly in the room outside $\Omega$. Of those, half were used for hyperparameter optimization, and half were used for the results 
presented here. Our experiments are performed for $200~\mathrm{Hz}$, 
$400~\mathrm{Hz}$, and $600~\mathrm{Hz}$.

\noindent\textbf{Network Architecture.}
For all experiments, we use MLPs with 4 hidden layers 
and 256 neurons per layer, with $\tanh$ activation. The RFF 
matrix $\bm B \in \mathbb{R}^{128\times3}$ is sampled from a Gaussian distribution with unit variance.

\noindent\textbf{Training.} We train each instance for $5\times 10^5$ iterations 
with the AdamW \cite{Loshchilov:ICLR2019} optimizer with an initial learning rate 
of $10^{-3}$. 
Every $10^4$ iterations, we decrease the learning rate by $10\%$. We employ a 
data loss weight $\lambda_{\text{data}} = 10^{-1}$ and a PDE loss weight of 
$\lambda_{\text{PDE}} = 10^{-3}$. Those weights have been optimised to achieve 
the lowest possible data loss at the expense of a slightly increased physics loss.

\subsection{Results}
We compare three different approaches. Our \textit{Baseline} is simple nearest 
neighbour interpolation~\cite{Murphy:ML}. 
We compare against interpolation with a NF without physics loss (\textit{NF}), 
which is trained solely with the data loss \eqref{eq:data_loss}. This approach is 
only implicitly regularized through the biases inherent to the network 
architecture~\cite{Ulyanov:CVPR2018}. 
Our proposed approach is interpolation by the phase-retrieval-based PINN 
(\textit{PRB-PINN}).

\noindent\textbf{Reconstruction Quality.} We compare the reconstruction quality 
of the different methods in Fig. \ref{fig:data_loss}. 
Across all frequencies and numbers of sensors, there is a consistent ordering 
of methods: Baseline performs worst, followed 
by NF, and our PRB-PINN performs best. 

As expected, reconstruction quality increases with the number of measurements. 
Furthermore, the interpolation problem becomes harder with increasing frequency 
as the sound field varies more rapidly in space. This leads to a lower 
reconstruction quality for all considered methods.

\begin{figure}[t!]
\centering
\begin{tabular}{cc|c}
\multicolumn{2}{c}{\footnotesize Ground truth} & \footnotesize Baseline
\\[1mm]
\includegraphics[height=26mm]{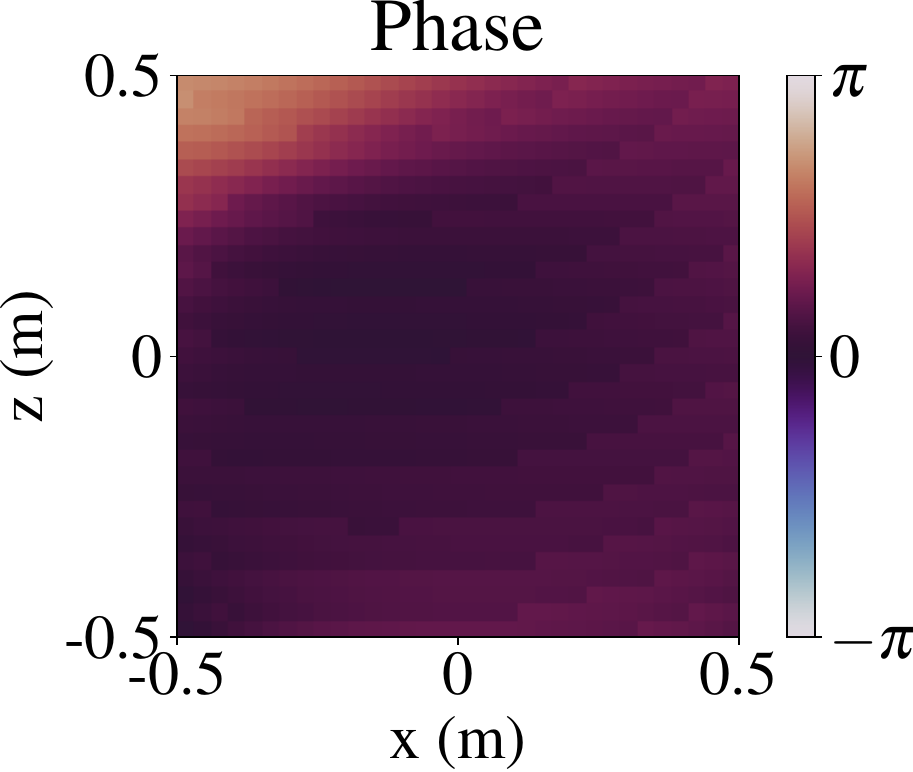}
\hspace{-3mm}
 &
\includegraphics[height=26mm]{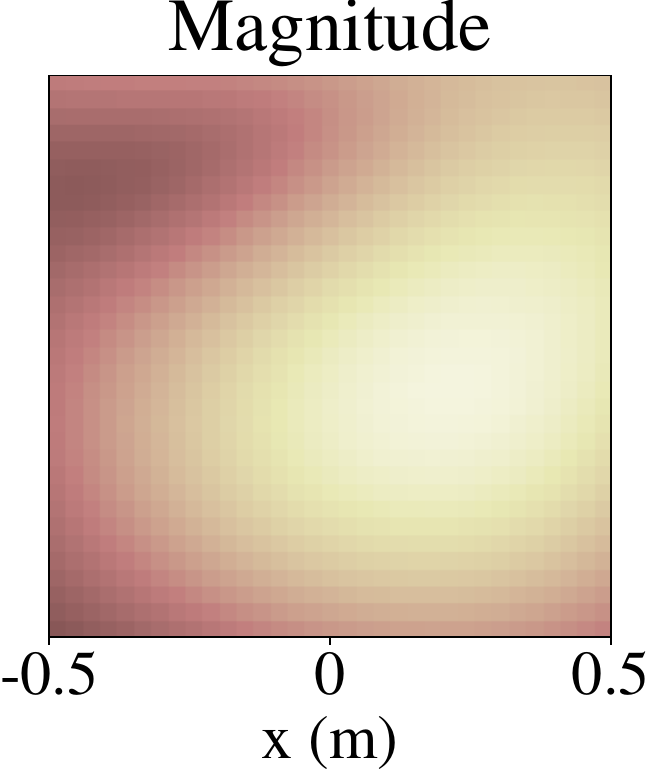}
 &
\includegraphics[height=26mm]{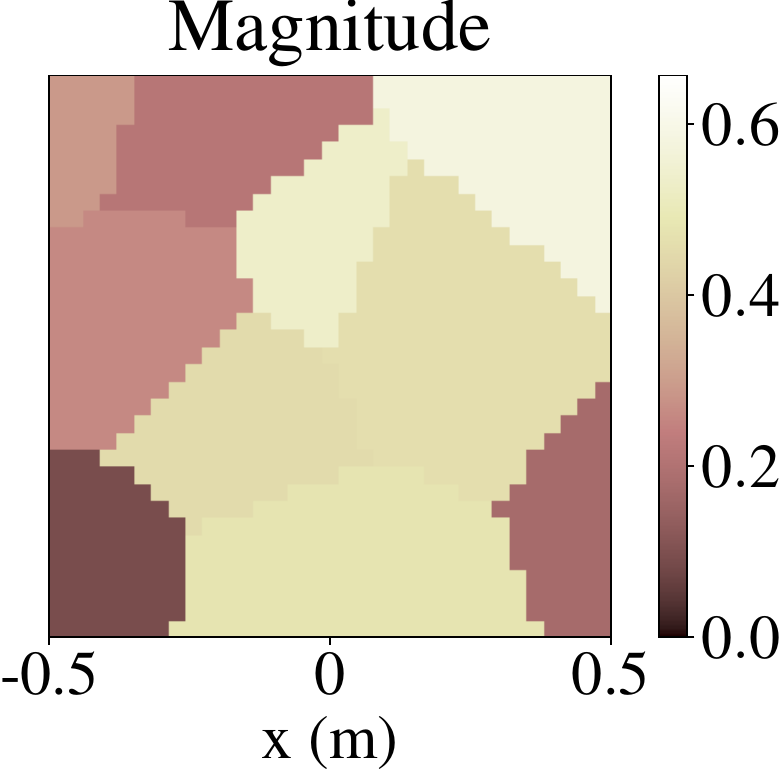}
 \\
\multicolumn{2}{c}{\footnotesize PRB-PINN} & \footnotesize NF\\
\includegraphics[height=26mm]{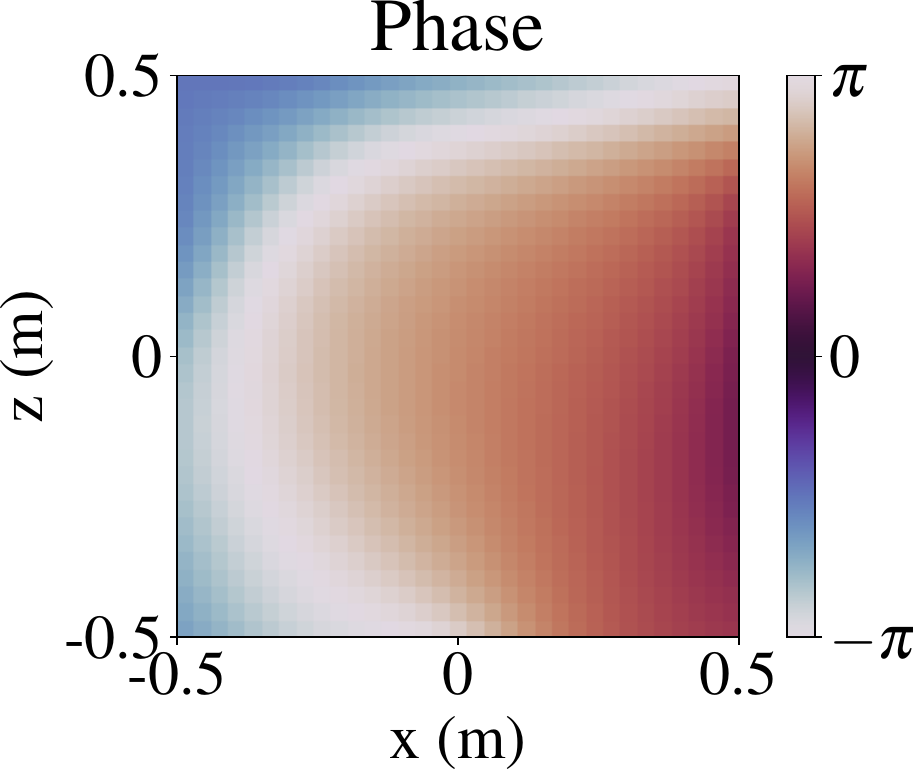}
\hspace{-3mm}
 &
\includegraphics[height=26mm]{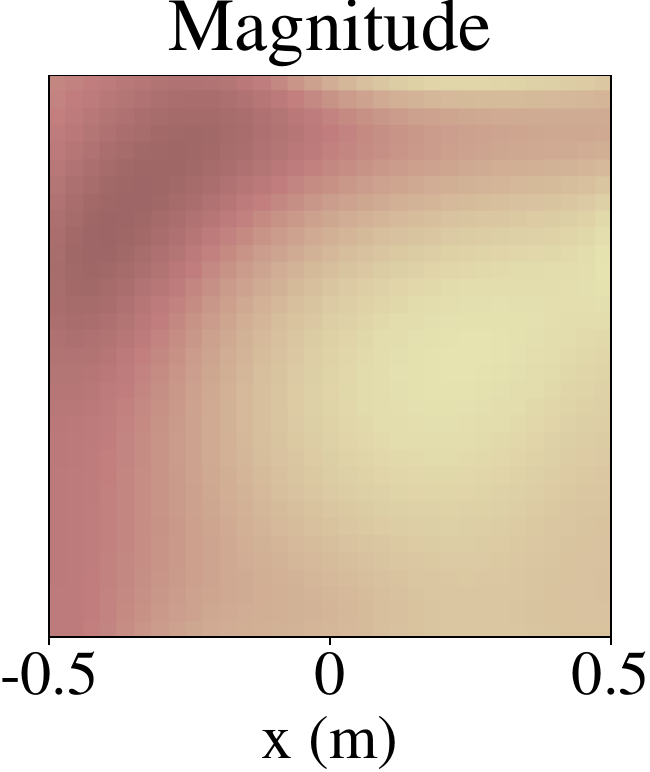}
 &
\includegraphics[height=26mm]{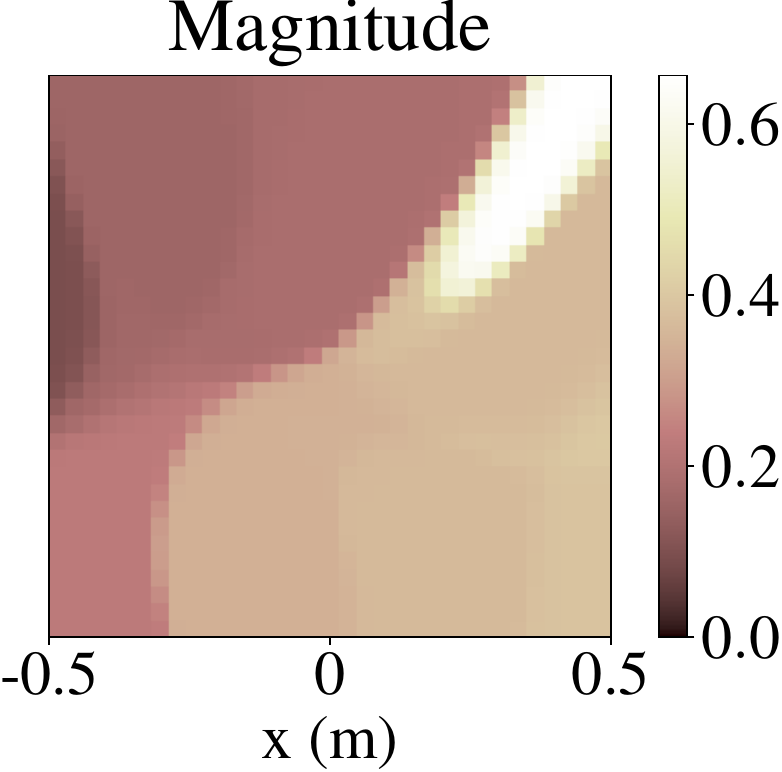}
\end{tabular}

\vspace{2mm}
{\footnotesize (a) $200$~Hz, number of measurements 20.}
\vspace{2mm}

\begin{tabular}{cc|c}
\hline\\[-2mm]
\multicolumn{2}{c}{\footnotesize Ground truth} & \footnotesize Baseline
\\[1mm]
\includegraphics[height=26mm]{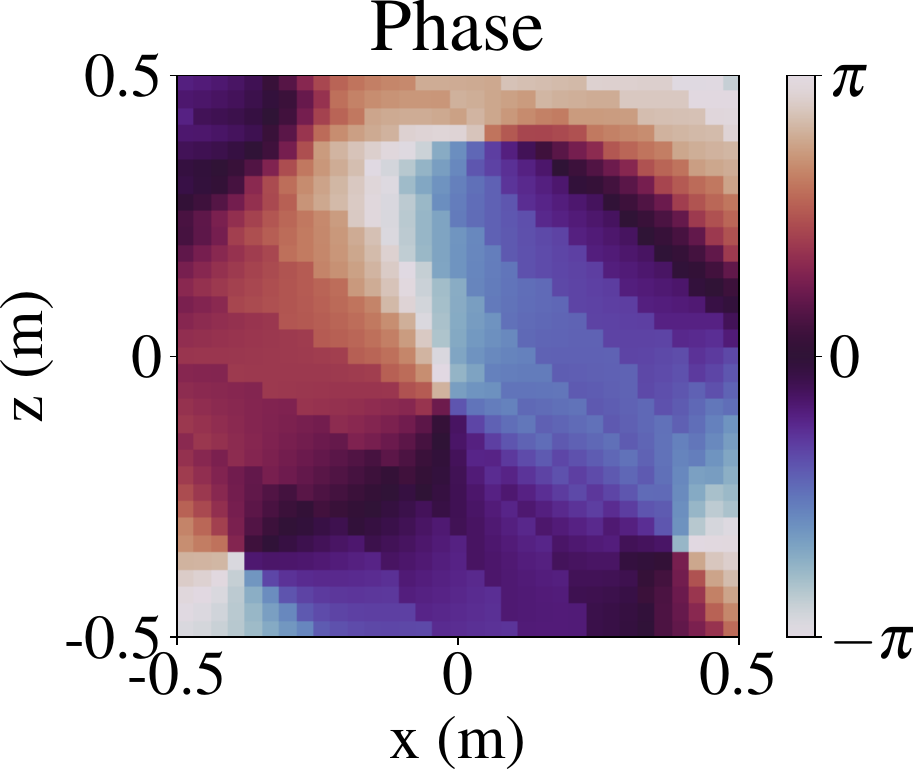}
\hspace{-3mm}
 &
\includegraphics[height=26mm]{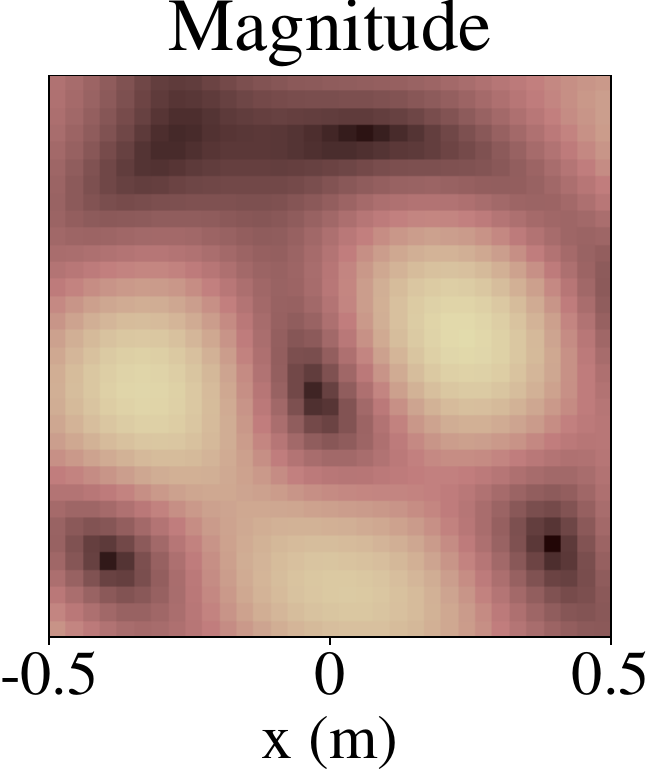}
 &
\includegraphics[height=26mm]{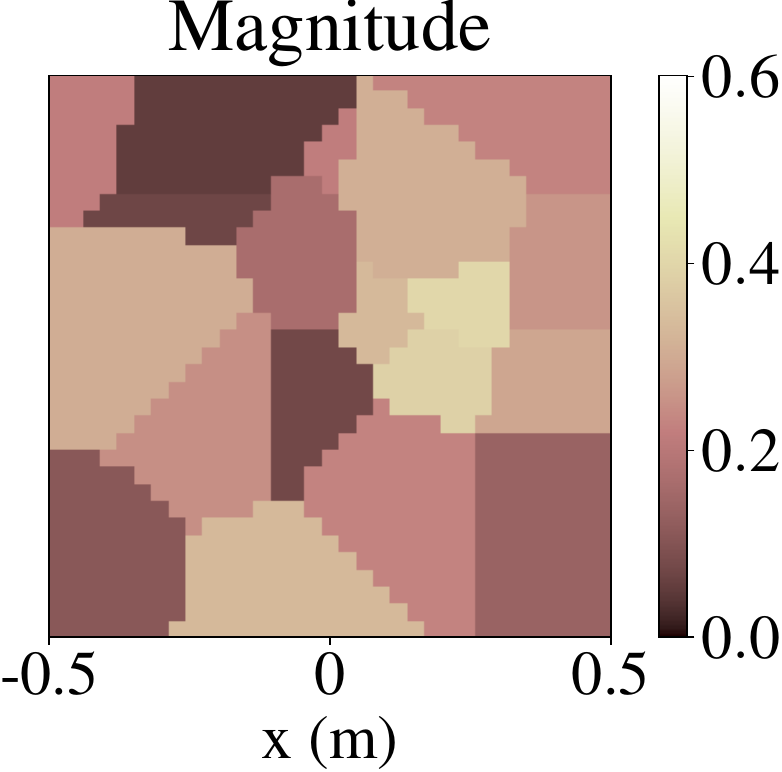}
 \\
\multicolumn{2}{c}{\footnotesize PRB-PINN} & \footnotesize NF\\
\includegraphics[height=26mm]{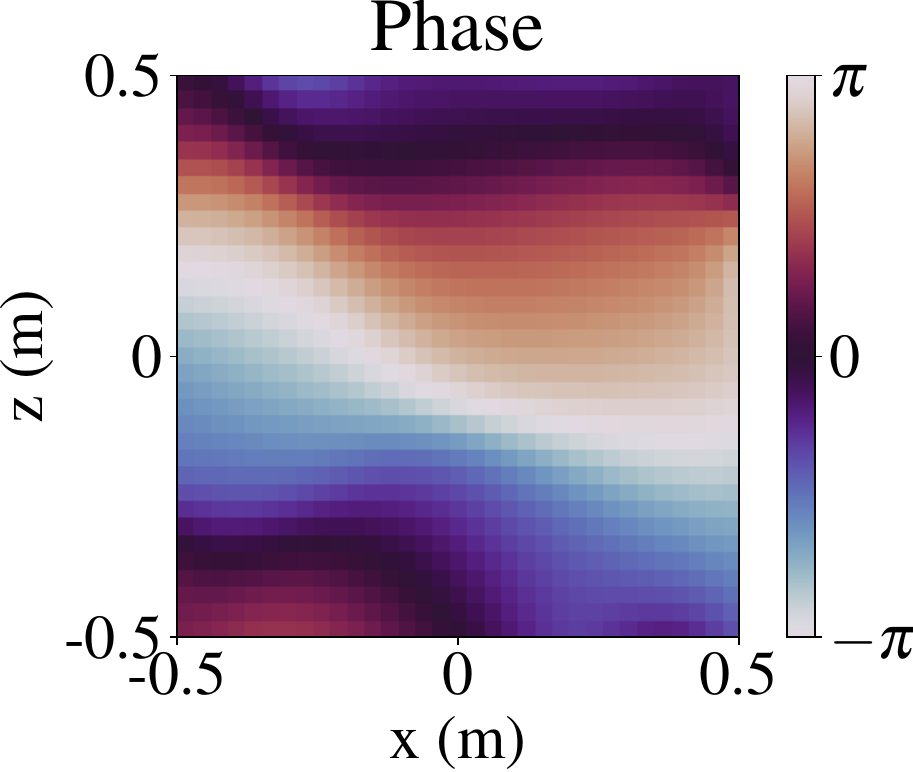}
\hspace{-3mm}
 &
\includegraphics[height=26mm]{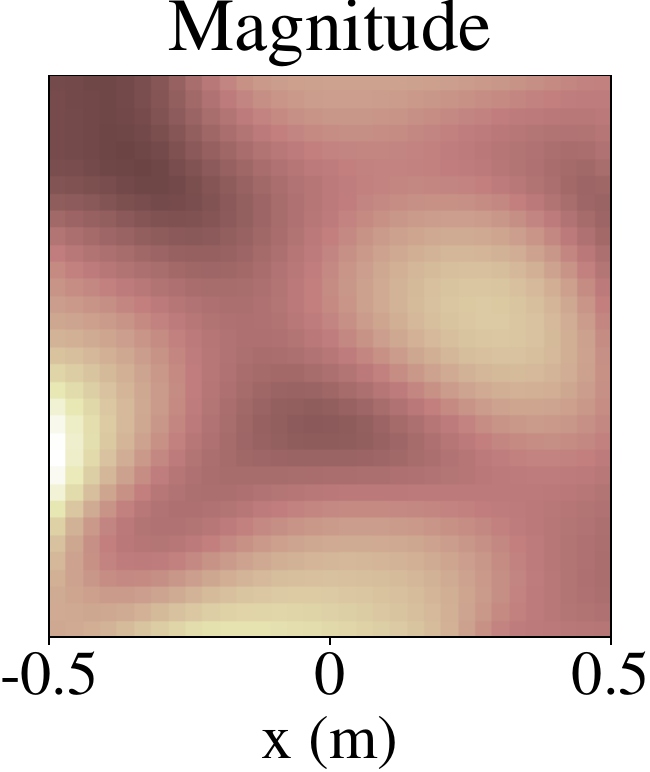}
 &
\includegraphics[height=26mm]{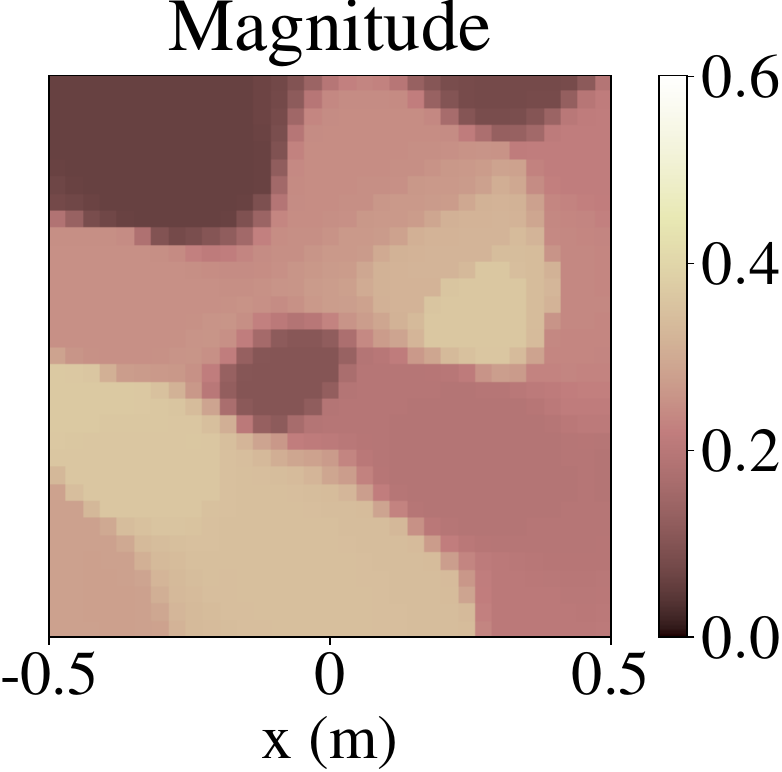}
\end{tabular}

\vspace{2mm}
{\footnotesize (b) $400$~Hz, number of measurements 50.}
\caption{Visualization of the magnitude distribution in the $x$-$z$ plane at 
$200$~Hz and $400$ Hz.}
\label{fig:visualisation}
\end{figure}

\noindent\textbf{Visual Evaluation.}
In Fig.~\ref{fig:visualisation}, we visualize the magnitude distributions in the 
$x$-$z$-plane of our target domain for different frequencies. Notably, we can 
see that the phase produced by the NN does not match the ground truth phase. 
However, it still leads to spatial variation with the correct frequency in the 
magnitude distribution, significantly increasing reconstruction quality.
Especially at a low frequency of $200~\mathrm{Hz}$, due to the low amount of 
variation in space, this reconstruction then matches the ground truth well. For 
higher frequencies, the space of physically plausible phases increases as well. 
As such, there can be a greater mismatch between reconstruction and ground truth.
%

\noindent\textbf{PDE Loss.} 
\begin{figure}[tbp]
\centering
\includegraphics[width=0.75\linewidth]{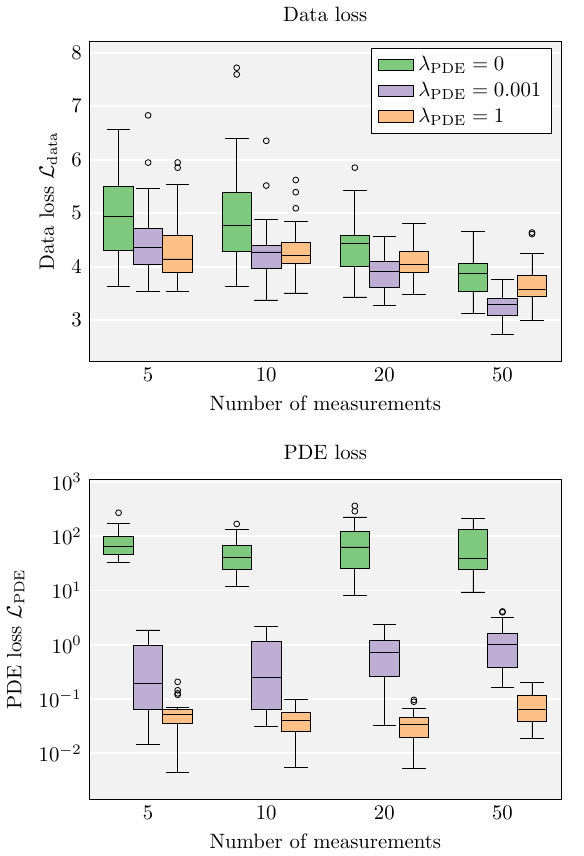}
\caption{Visualisation of the impact of the PDE loss weight 
$\lambda_{\text{PDE}}$. A PDE loss either lower or higher than the chosen one 
results in worse reconstruction quality. }
\label{fig:physics}
\end{figure}
Fig.~\ref{fig:physics} explores the influence of the PDE loss weight 
$\lambda_{\text{PDE}}$. Especially for higher numbers of measurement locations $M$, 
increasing the weight too far can lead to a higher test loss, even though the 
physics loss still decreases.
This highlights the importance of adequate loss balancing for training PINNs.

\section{Conclusion}
We proposed the phase-retrieval-based PINN for the magnitude field estimation in 
the frequency domain. Due to the inaccessible phase data, current methods for the 
magnitude field estimation rely only on the measured magnitude data and do not 
employ the governing PDE that the original complex-valued function of the 
estimation target should satisfy. We proposed the network architecture for 
estimating magnitude and phase distributions using NF, and penalized the 
deviation of the reconstructed complex-valued function from the Helmholtz 
equation. We demonstrated that our phase-retrieval PDE loss is effective by 
comparing the proposed method with the NF without the PDE loss in the 
experiments. 


\bibliographystyle{IEEEbib_mod}
\bibliography{str_def_abrv,skoyamalab_en,refs}

\end{document}